\documentclass[11 pt]{article}

\usepackage{amsmath,amstext,amsgen,amsbsy,amsopn,amsfonts,graphicx,
theorem,amssymb,shadow}

{\theorembodyfont{\rmfamily} \theoremheaderfont{\itshape} 
}

\begin{document}

\title{Ground Control to Niels Bohr: Exploring Outer Space with Atomic 
Physics}

\author{Mason A. Porter and Predrag Cvitanovi\'c\footnote{Mason Porter
    is a Visiting Assistant Professor in the School of Mathematics and
    a Research Associate Member of the Center for Nonlinear Science at
    Georgia Institute of Technology.  Predrag Cvitanovi\'c is the Glen
    Robinson Chair in Nonlinear Science in the School of Physics at
    Georgia Institute of Technology.  The authors acknowledge Shane
    Ross and Turgay Uzer for useful discussions and critical readings
    of this paper.  They also thank Jerry Marsden and Thomas Bartsch
    for providing them with their preprints.  MAP also acknowledges
    support provided by an NSF VIGRE grant awarded to the School of
    Mathematics at Georgia Tech.}}

\maketitle

\section*{To the Sun and Back}

\begin{figure}
        \begin{centering}
                \includegraphics[width = 0.9\textwidth]
{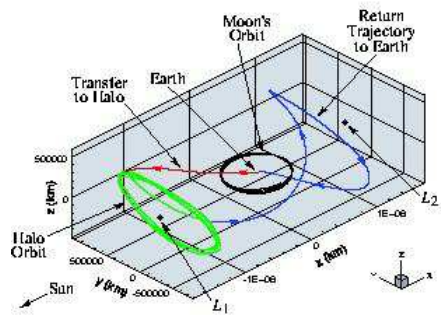}


                \caption{Planned trajectory for the Genesis spacecraft, whose 
                  several-year mission is to collect charged particles
                  from the solar wind and return them to Earth.  The
                  trajectory was chosen to take Genesis sufficiently
                  far away from Earth's geomagnetic field so that
                  solar wind samples could be collected before
                  interacting with that field.  It reached the first
                  Lagrange point ($L_1$) on 16 November 2001, setting
                  up five halo loops about $L_1$ (lasting 30 months)
                  that began the scientific portion of the mission.
                  Sample collection lasted from 3 December 2001 until
                  1 April 2004.  Genesis released its sample return
                  capsule on 8 September 2004 (which arrived on 4
                  October) and then headed back to $L_1$, which it was
                  scheduled to leave in February 2005, after which it
                  was slated to begin orbiting around the Sun just
                  inside Earth's orbit.  (Figure courtesy Roby Wilson,
                  Jet Propulsion Laboratory/California Institute of
                  Technology.)}
                \label{mission}
        \end{centering}
\end{figure}

8 August 2001 was an exciting day for scientists studying nonlinear
dynamics.  With a trajectory designed using techniques from the theory
of dynamical systems, NASA launched the spacecraft Genesis towards the
Sun to collect pieces of it (called solar wind).  When Genesis
completes its mission (see Fig. \ref{mission}), scientists may
determine not only the composition of the Sun but also whether Earth
and the other planets have the same constituents.  The samples
collected by the Genesis mission of NASA's Discovery program will be
studied extensively for many years now that the spacecraft has
returned some of its souvenirs to Earth.  A sample return capsule,
containing the first extraterrestrial matter returned by a U.S.
spacecraft since 1972, was released by Genesis on 8 September 2004 and
arrived at the Johnson Space Center in Houston, TX on 4 October.  It
was subsequently announced in March 2005 that ions of Solar origin
were indeed present in one of the wafer fragments \cite{nasa,nasa2}.

M. Lo of the Jet Propulsion Laboratory, who led the development of the
Genesis mission design, worked with Caltech mathematician J. Marsden,
Georgia Tech physicist T. Uzer, and West Virginia University chemist
C. Jaff\'e on the statistical analysis of transport phenomena.  Why?
The Genesis trajectory constitutes a highly unstable orbit (controlled
by the Lagrange equilibrium points) of the infamous celestial three
body problem studied by H. Poincar\'e and others.  Some of the most
dangerous near-earth asteroids and comets follow similar chaotic
paths, which have the notorious property that they can be resolved
with numerical simulations only up to some finite time.

In a turn of events that would have astonished anyone but N. Bohr, we
now know that chaotic trajectories identical to those that govern the
motions of comets, asteroids, and spacecraft are traversed on the
atomic scale by highly excited Rydberg electrons
\cite{jaff1,jaff2,jaff3,jaff4,nonlin}.  This almost perfect parallel
between the governing equations of atomic physics and celestial
mechanics implies that the transport mechanism for these two
situations is virtually identical: On the celestial scale, transport
takes a spacecraft from one Lagrange point to another until it reaches
its desired destination. On the atomic scale, the same type of
trajectory transports an electron initially trapped near the atom
across an escape threshold (in chemical parlance, across a
``transition state''), never to return.  The orbits used to design
space missions thus also determine the ionization rates of atoms and
chemical-reaction rates of molecules!

Recent work \cite{nonlin,waa1,waa2} also offers hope that researchers
may eventually overcome one of the current outstanding challenges of
nonlinear science: how does one describe chaotic dynamics in systems
with many degrees-of-freedom but still too few to be amenable to the
methods of statistical physics?  The concept of ``chaos'' is
well-understood only for low-dimensional systems, as few methods deal
successfully with higher-dimensional dynamics.  Transition state
theory is one such tool.

The large-scale chaos present in the Solar System is weak enough that the 
motion of most planets appears regular on human time scales.  Nevertheless, 
small celestial bodies such as asteroids, comets, and spacecraft can behave 
in a strongly chaotic manner, and it is important to be able to predict the 
behavior of populations of these smaller celestial bodies not only to design
 gravitationally-assisted transport of spacecraft but also to develop a 
statistical description of populations of comets, near-Earth asteroids, and 
zodiacal and circumplanetary dust \cite{jaff3}.

This is precisely the challenge faced by atomic physicists and
chemists in computing ionization rates of atoms and molecules.  In
brute force approaches, this is accomplished via large numerical
simulations that track the orbits of myriad test particles with as many
interactions as desired.  In practice, however, such techniques are
computationally intensive and convey little insight into a system's
key dynamical mechanisms.  A theoretically grounded approach relies on
transition state theory \cite{jaff3}.  ``Transition states'' are
surfaces (manifolds) in the many-dimensional phase space (the set of
all possible positions and momenta that particles can attain) that
regulate mass transport through bottlenecks in that phase space; the
transition rates are then computed using a statistical approach
developed in chemical dynamics \cite{nonlin}.  In such analyses, one
assumes that the rate of intramolecular energy redistribution is fast
relative to the reaction rate, which can then be expressed as the
ratio of the flux across the transition state divided by the total
volume of phase space associated with the reactants.

In the next few sections, we'll delve a bit deeper into this story.
We start with an introduction to transition state theory and then show
how this theory from atomic and molecular physics can be used on the
much grander celestial scale.  We then close with some recent
extensions and a brief summary.

\section*{Back in the Saddle Again}

Before heading off into outer space, we need to examine things on a
much smaller scale---namely, simple chemical reactions between ions
and small molecules.

{\it Transition state theory} has its origins in early 20th century
studies of the dynamics of chemical reactions.  Consider, for example,
the collinear reaction between the hydrogen atom $H$ and the hydrogen
molecule $H_2$ in which one hydrogen atom switches partners.  In the
1930s, Eyring and Polanyi \cite{eyring} studied this chemical
reaction, providing the first calculation of the potential energy
surface of a reaction.  This surface contains a minimum associated to
the reactants and another minimum for the products; they are separated
by a barrier that needs to be crossed for the chemical reaction to
occur.  Eyring and Polanyi defined the surface's ``transition state''
as the path of steepest ascent from the barrier's saddle point.  Once
crossed, this ``transition state'' could never be recrossed.

The notion of a transition state as a ``surface of no return'' defined
in coordinate space was immediately recognized as fundamentally
flawed, as recrossing can arise from dynamical effects due to coupling
terms in the kinetic energy.  (See Ref.~\cite{jaff1} for further
historical details.)  Pechukas demonstrated that the surface of
minimum flux, corresponding to the transition state, must be an
unstable periodic orbit whose projection onto coordinate space
connects the two branches of the relevant equipotentials
\cite{pechukas}.  As a result, these surfaces of minimum flux are
called ``periodic orbit dividing surfaces'' or PODS.

\begin{figure}
        \begin{centering}
                \includegraphics[width = .95\textwidth]
{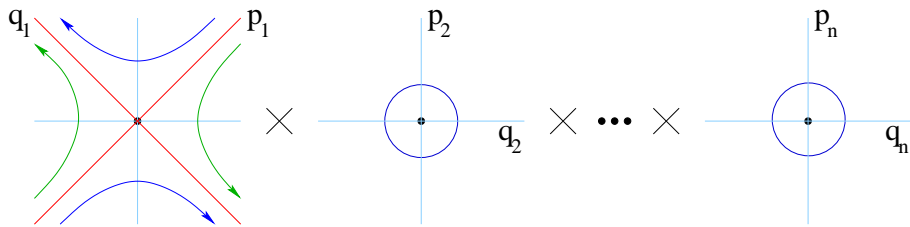}
                \caption{The linearized dynamics of a transition state.  One 
                  degree-of-freedom yields a pair of real eigenvalues
                  of opposite sign (shown by the saddle projection on
                  the left), and the others give pure imaginary
                  conjugate pairs of eigenvalues (indicated by the
                  center projections) The blue trajectories in the
                  left panel are reactive, whereas the green ones are
                  not.}
                \label{pod}
        \end{centering}
\end{figure}


Despite the specificity of the $H_2 + H \rightleftharpoons H + H_2$
reaction, a transition state is a very general property of Hamiltonian
dynamical systems describing how a set of ``reactants'' evolves into a
set of ``products'' \cite{wigner2}.  Transition state theory can be
used to study ``reaction rates'' in a diverse array of physical
situations, including atom ionization, cluster rearrangement,
conductance through microjunctions, diffusion jumps in solids, and (as
we shall discuss) celestial phenomena such as asteroid escape rates
\cite{jaff1,jaff2,jaff3,jaff4}.

E. Wigner recognized very early that in order to develop a rigorous
theory of transition states, one must extend the notions above from
configuration space to the phase space of positions and momenta
\cite{jaff4,wigner1}.  (Each position-momentum pair constitutes one of
the system's ``degrees-of-freedom'' [DOF].)  The partitioning of phase
space into separate regions corresponding to reactants and products
thereby becomes the theory's goal, progress towards which has required
advances in both dynamical systems theory and computational hardware.

For two DOF Hamiltonian systems, the stable and unstable manifolds of
the orbit discussed provide an invariant partition of the system's
energy shell into reactive and nonreactive dynamics.  The defining
periodic orbit also bounds a surface in the energy shell (at which the
Hamiltonian is constant), partitioning it into reactant and product
regions.  This, then, defines a surface of no return and yields an
unambiguous measure of the flux between reactants and products.  In
systems with three or more DOF, however, periodic orbits and their
associated stable and unstable manifolds do not partition energy
shells (their dimensionality is insufficient) \cite{lich}, so one
needs to search instead for higher-dimensional analogs of PODS
\cite{jaff4}.

Consider an $n$ DOF Hamiltonian system with an equilibrium point, the
linearization about which has eigenvalues $\pm \lambda$, $\pm
i\omega_j$, $j \in \{2\,,\dots\,,n\}$\,, where $\lambda\,, \omega_j \in
\mathbb{R}$.  That is, we are considering situations in which the
stable and unstable manifolds are each one-dimensional.  (There exist
chemical reactions with higher-dimensional stable and unstable
manifolds, but theoretical chemists do not really know how to deal
with them yet.)  Also assume that the submatrix corresponding to the
imaginary eigenvalues is symmetric, so that its complexification is
diagonal.  One can then show that in the vicinity of the saddle point,
the normal form of this Hamiltonian is \cite{gucken}
\begin{align}
  H &= \lambda q_1p_1 + \sum_{i = 2}^{n}\frac{\omega_i}{2}(p_i^2 + q_i^2) 
+ f_1(q_2,\dots,q_{n},p_2,\dots,p_{n},I) \notag \\
  &\quad + f_2(q_2,\dots,q_{n},p_2,\dots,p_{n})\,, \label{ham}
\end{align}
where $(q_1,\dots,q_{n},p_1,\dots,p_{n})$ are the canonical
coordinates, $I := q_1p_1$, and the functions $f_1$ and $f_2$ are at
least third order and account for all the nonlinear terms in
Hamilton's equations.  Additionally,
$f_1(q_2,\dots,q_{n},p_2,\dots,p_{n},I) = 0$ when $I = 0$.
Although (\ref{ham}) is constructed locally, it continues to hold as
parameters are adjusted until a bifurcation occurs.

The simplest example is the linear dynamical system with Hamiltonian
\begin{equation}
  H = \frac{1}{2}p_\xi^2 - \frac{\kappa^2}{2}\xi^2 
+ \frac{1}{2}\sum_{i = 2}^{n}(p_i^2 + \omega_i^2q_i^2) 
\end{equation}
consisting of $n - 1$ decoupled linear (``harmonic'') oscillators and
a decoupled saddle point, which can be obtained from the linearization
of (\ref{ham}) by a rotation in phase space (see Fig.~\ref{pod}).  The
first DOF $(\xi,p_\xi)$ gives the ``reaction coordinates'' and the
other $n - 1$ DOF are ``bath coordinates.''  A trajectory is called
``reacting'' if $\xi$ changes sign as one traverses it.

Such considerations can be generalized from this linear situation to
the fully nonlinear Hamiltonian (\ref{ham}) needed to describe
chemical reactions by considering higher-dimensional analogs of saddle
points called {\it normally hyperbolic invariant manifolds} (NHIMs)
\cite{jaff4,wig}.  The descriptor `normally hyperbolic' means that in
the linearization of (\ref{ham}), the growth and decay rates of the
dynamics normal to the NHIM (constituting the ``reaction'') dominate
the growth and decay rates of the dynamics tangent to the NHIM, which
is obtained as follows: The dynamics of (\ref{ham}) are described by
the $(2n-1)$-dimensional energy surface $H = \mbox{constant} > 0$.  If
$p_1 = q_1 = 0$, it follows that $\dot{q}_1 = \dot{p}_1 = 0$, which
yields a $(2n-2)$-dimensional invariant manifold, whose intersection
with the energy surface gives the NHIM.  The $(q_1,p_1)$ coordinates
describe the directions normal to the NHIM.  Additionally, NHIMs
persist under perturbations, so one can transform back from
(\ref{ham}) to the original Hamiltonian system derived by physical or
chemical considerations.  The stable and unstable manifolds of the
NHIM are known explicitly and act as impenetrable (invariant)
boundaries between reactive and nonreactive trajectories \cite{jaff4}.

Before proceeding, let's consider the example of hydrogen ionization
in crossed electric and magnetic fields, as described by the
Hamiltonian
\begin{align}
  H(x_1,x_2,x_3,P_1,P_2,P_3) &= \frac{1}{2}(P_1^2 + P_2^2 + P_3^2) 
- \frac{1}{R} \notag \\ &\quad + \left[\frac{1}{2}(x_1P_2 - x_2P_1) 
+ \frac{1}{8}(x_1^2 + x_2^2) - \epsilon x_1 - \sqrt{\epsilon}\right]\,, 
\label{cross}
\end{align}
where $R = \sqrt{(x_1 + \sqrt{\epsilon})^2 + x_2^2 + x_3^3}$.  The
equilibrium at the origin has two imaginary pairs of eigenvalues and
one real pair, so it's a center-saddle-center.  The Hamiltonian
(\ref{cross}) can be transformed to its normal form, whose lowest
order term is
\begin{equation}
  H_2 = \mu x_1P_1 + \frac{\omega_1}{2}(x_2^2 + P_2^2) 
+ \frac{\omega_2}{2}(x_3^2 + P_3^2)\,.
\end{equation}
As required, the saddle variables $(x_1,P_1)$ appear only in the
combination $x_1P_1$, so a NHIM can be constructed as discussed above
and one can easily study which trajectories react and which do not.

\section*{Hitchhiking the Solar System with Bohr and Poincar\'e}

Volume 7 (1885-86) of {\it Acta Mathematica} included the announcement
that King Oscar II of Sweden and Norway would award a medal and 2500
kroner prize to the first person to obtain a global general solution
to the $n$-body celestial problem \cite{diacu}.  Henri Poincar\'e,
then thirty-one years old, had long been fascinated with celestial
mechanics.  His first paper, published in 1883, treated some special
solutions of the 3-body problem.  The following year, Poincar\'e
published a second paper on the topic, but he had not touched
celestial mechanics since then.  Nevertheless, he had developed new
qualitative techniques for studying differential equations that he
felt would provide a good intuitive basis for his attempt to solve the
$n$-body problem.

In the treatise that resulted from his attempt to win King Oscar II's
prize \cite{poinorig1,poinorig2,poinorig3}, Poincar\'e laid the
foundations for dynamical systems theory, developing integral
invariants to prove his recurrence theorem, a new approach to periodic
solutions and stability, and much more.  Some of his results clashed
with his prior intuition, and there were others that he felt were true
but that he was unable to establish rigorously (the world would have
to wait for the likes of G. Birkhoff, S. Smale, and others).  After
more than two years of working on the $n$-body problem, the solution
began to take shape.  One of the problem's secrets was revealed by the
3-body problem: Poincar\'e proved that there did not exist uniform
first integrals other than $H = \mbox{constant}$, so that even the
3-body problem could not be ``integrated.''  Chaos was here to stay!

Now that we have discussed the mathematics of transition states, let's
see how they can help us not only on atomic problems but also on
celestial ones.  To do this, we will use the old adage that the same
equations have the same solutions: Namely, a suitable coordinate
change transforms the Hamiltonian describing the celestial restricted
three body problem (RTBP) into the Hamiltonian (\ref{cross})
describing hydrogen ionization in crossed electric and magnetic fields
\cite{jaff3}.  The term ``restricted'' is used when the mass of one
body is assumed to be so small that it does not influence the motion
of the other two bodies, which follow circular orbits around their
center of mass.  It is also assumed that all three orbits lie in a
common plane \cite{diacu,koon}.

In conventional coordinates, the RTBP is described by the Hamiltonian
\begin{equation}
  H = \frac{1}{2}(p_x^2 + p_y^2) - (xp_y - yp_x) - \frac{1-\mu}{r_1} 
- \frac{\mu}{r_2} - \frac{1}{2}\mu(1-\mu) = E\,,
\end{equation}
where $E$ is the energy, $r_1 = \sqrt{(x+\mu)^2 + y^2}$, $r_2 =
\sqrt{(x - 1 + \mu)^2 + y^2}$, and the masses of the bodies are $m_s =
1 - \mu$ and $m_p = \mu < m_s$.  (The notation is chosen so that one
thinks of $m_s$ as the Sun's mass and $m_p$ as a planet's mass.)  The
coordinate system rotates with the period of Jupiter about the
Sun-Jupiter center of mass.  The Sun and Jupiter are located
respectively, at $(-\mu,0)$ and $(1-\mu,0)$.  The position of the the
third body (say, an asteroid) relative to the Sun and the planet is
$(x,y)$.

\begin{figure}
        \begin{centering}
                \includegraphics[width = 0.95\textwidth]{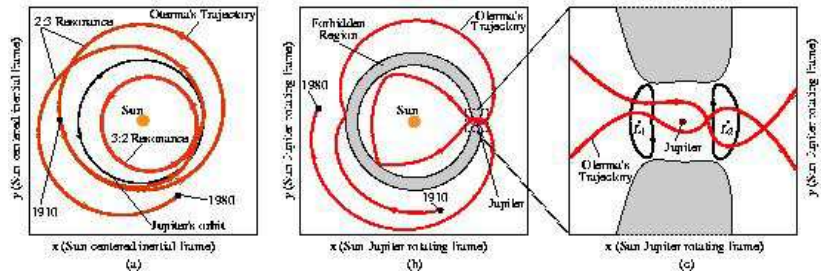}
                \caption{Resonance transition of the Jovian comet Oterma. (a) 
                  The dynamics in heliocentric coordinates. (b) The
                  dynamics in a reference frame rotating with Jupiter.
                  (c) Magnified view of the bottleneck region from
                  (b).  (Figure courtesy Shane D. Ross, University of
                  Southern California.)}
                \label{oterma}
        \end{centering}
\end{figure}

An example is provided by Jupiter's comets such as {\it Oterma} which
shuttle back and forth between complex heliocentric orbits lying,
respectively, interior and exterior to Jupiter's orbit \cite{jaff3}
(see Fig.~\ref{oterma}).  (Oterma lies in the same energy regime as
Shumaker-Levy 9, so it is destined to one day crash into Jupiter.)
Jupiter often temporarily `captures' such comets while they make these
transitions.  The interior orbits are generally near a 3:2 resonance,
with Oterma making three revolutions about the Sun for every two solar
revolutions of Jupiter (in the inertial frame), whereas the exterior
ones are near a 2:3 resonance.  In a frame rotating with Jupiter, the
transition between resonances occurs in a ``bottleneck'' region in
configuration space.

The celebrated ``Jacobi integral'' (a constant of motion) provides a
dynamical invariant that divides phase space into reactant (interior)
and product (exterior) regions, which are separated by the narrow
bottleneck containing Jupiter and two of the Lagrange points, $L_1$
and $L_2$.  The passage of celestial bodies like comets through the
bottleneck is then regulated by phase space structures near $L_1$ and
$L_2$, which are both saddle points.  The transition states in this
problem, controlling transport through the bottleneck and hence the
conversion of ``reactants'' to ``products,'' are the periodic orbits
around $L_1$ and $L_2$.  With these structures identified, C. Jaff\'e
{\it et al.}  have accurately computed average transport rates
(corresponding to asteroid escape rates) using
Rice-Ramsperger-Kassel-Marcus theory and checked the predicted rates
against large-scale numerical simulations \cite{jaff3}.

\section*{Meanwhile, Back on Earth...}

The story doesn't end with the work discussed in this note.  On the
practical side, discussions at NASA are currently underway about the
possibility of an extended Genesis mission that would keep the
spacecraft in the Earth-Moon system for the next several years
\cite{nasa}.

On the theoretical side, the mathematics, physics, and chemistry
communities remain hard at work.  Recent discoveries include a
computational procedure based on NHIMs to detect high-dimensional
chaotic saddles in three DOF Hamiltonian systems (and the application
of this technology to, for example, the three-dimensional Hill's
problem) \cite{waa1}, mathematical refinements of earlier
constructions of transition states \cite{waa2}, and the effect of
noise on transition states \cite{thomas}.  Current work on space
mission design includes the use of set-oriented methods and ideas from
graph theory to go beyond transition state theory \cite{design} and
the merging of tube dynamics with a Monte Carlo approach to examine
the invariant manifolds emanating from transition states
\cite{grabern}.

It is a time-honored scientific tradition that the same equations have
the same solutions.  When it comes to $3$-body problems, this implies
that the same chaotic trajectories that govern the motions of comets,
asteroids, and spacecraft are traversed on the atomic scale by highly
excited Rydberg electrons.  Such unanticipated connections between
microscopic and celestial phenomena are not only intellectually
gratifying but also have practical engineering applications in the
aerospace and chemical industries.  Moreover, the progress made would
hardly be conceivable without this particular mix of specialists
recruited by M. Lo.  Clearly, chemists, astronomers, and
mathematicians have much to discuss!

Additionally, while it is paramount in many problems to slay the
dragon of chaos so that order can reign, just the opposite is true
here---the goal is to create a big enough (chaotic) saddle and ride
this dragon on the (Normally Hyperbolic) Invariant Manifold
Superhighway!  The Genesis mission shows that chaos can, in fact, be
good.

\bibliographystyle{siam}

\end{document}